\documentclass[aip,jcp,showpacs]{revtex4}
\usepackage{lipsum}
\usepackage{bm}
\usepackage{epsfig,amsopn}
\usepackage{graphicx}
\usepackage{amsmath,amssymb}
\usepackage{natbib}
\usepackage{color}
\usepackage{latexsym}
\usepackage{amsfonts}
\usepackage{amssymb}
\usepackage{comment}

\newcommand{\bea}{\begin{eqnarray}}
\newcommand{\eea}{\end{eqnarray}}

\def\beq{\begin{eqnarray}}
\def\eeq{\end{eqnarray}}

\begin{document}
\title{ From dissipative dynamics to studies of heat transfer at the nanoscale}
\author{Nazim Boudjada$^{1,2}$}
\author{Dvira Segal$^2$}

\affiliation{$^1$Ecole Polytechnique de Montreal, Montreal, Quebec, Canada H3C 3A7}
\affiliation{$^2$Chemical Physics Theory Group, Department of Chemistry, University of Toronto,
80 Saint George St. Toronto, Ontario, Canada M5S 3H6}

\date{\today}


\begin{abstract}
We study in a unified manner the dissipative dynamics and the
transfer of heat in the two-bath spin-boson model.
We use the Bloch-Redfield (BR) formalism, valid in the very weak system-bath coupling limit, the noninteracting-blip
approximation (NIBA), applicable in the non-adiabatic limit,
and iterative,  numerically-exact path integral tools.
These methodologies were originally developed for the description of the dissipative dynamics of a
quantum system, and here they are applied to explore the problem
of quantum energy transport in a non-equilibrium setting.
Specifically, we study the weak-to-intermediate system-bath coupling
regime at high temperatures $k_BT/\hbar>\epsilon$, with $\epsilon$
as the characteristic frequency of the two-state system.
The BR formalism and NIBA can lead to close results for the dynamics of the
reduced density matrix (RDM) in a certain range of parameters. However, relatively
small deviations in the RDM dynamics propagate into significant qualitative discrepancies in the transport behavior.
Similarly, beyond the strict non-adiabatic limit NIBA's prediction for the heat current is qualitatively
incorrect: It fails to capture the turnover behavior of the current with
tunneling energy and temperature.
Thus, techniques that proved meaningful for describing the RDM dynamics, to some extent even
beyond their rigorous range of validity,
should be used with great caution
in heat transfer calculations, since qualitative-serious failures
develop once parameters are mildly stretched beyond the techniques' working assumptions.
\end{abstract}



\maketitle

\section{Introduction}

Quantum impurity models, comprising a subsystem in an environment, embody complex processes in condensed
phases: electron and exciton transfer in solids, solutions, glasses
and biomolecules \cite{weiss,nitzan}, screening of a magnetic impurity by the Fermi sea electrons
 \cite{legget}, electronic conduction of
molecules \cite{nitzan}, and the decoherence behavior of superconducting qubits \cite{legget,karyn}.
The dissipative dynamics of impurity models has been explored intensively by time-evolving the subsystem
(reduced) density matrix, revealing mechanisms of decoherence and
relaxation towards equilibrium.
Here, as a case study, we focus on the spin-boson (SB) model
with a two-level system (TLS) immersed in a bath of harmonic
oscillators \cite{weiss}.

\begin{figure}[htbp]
\vspace{-2mm} \hspace{0mm}
{\hbox{\epsfxsize=50mm \epsffile{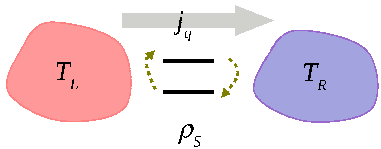}}}
\vspace{-2mm}
\caption{Scheme of the non-equilibrium spin-boson model, a minimal picture
for studying heat transport in anharmonic nanojunctions.
In this work we simulate both the dynamics of the (spin) reduced density matrix
$\rho_S$ and the heat current behavior $j_q$ using approximate methods and numerically exact simulation tools.
}
\label{scheme}
\end{figure}

Beyond the question of decoherence and dissipation,
impurity models can be employed for exploring fundamentals of quantum
transport and quantum thermodynamics, when placing the subsystem between two
reservoirs maintained e.g. at different chemical potentials or
temperatures. In this scenario, the reservoirs exchange charge, spin,
or energy carriers through the subsystem, with
quantities of interest as (charge, energy,
spin) currents in the system, as well as  high order cumulants of currents.
In the context of nanoscale heat transfer and phononics \cite{dhar08,wangrev,lirev12},
the ``non-equilibrium spin-boson model" (NESB),  
with a TLS bridging two thermal reservoirs at different temperatures, has been
suggested as a toy model for studying the phenomenology of quantum
heat transfer in anharmonic junctions \cite{sn05}, see Fig. \ref{scheme}
for a schematic representation.


The dissipative dynamics of the SB model has been examined
systematically by comparing predictions from different techniques.
Results have been organized in several reviews
\cite{weiss,legget}, and the problem still provides an active area for exploration,
see for example Refs.
\cite{thoss01,delft,thorwart,reichman11}. In contrast, the analysis of heat
transfer characteristics in the corresponding NESB model is a
relatively new problem and a systematic comparison of
results from different techniques is still missing.
One should note that the computation of transport characteristics in
the NESB model (and other non-equilibrium impurity models)
relies on a nontrivial extension of
open quantum systems methodologies: To calculate the current one needs to follow the
dynamics of other operators beyond the reduced density matrix
(RDM): two-time correlation functions of subsystem's operators or
expectation values of bath operators.
The thermal properties of the NESB nanojunction have been analyzed on the basis of perturbative quantum
master equations \cite{sn05,segal06,claire09,ro11,thingna12,redfield},
Keldysh Green's function expansions, \cite{vtw10,neGF,Aviv,wu14}, and
the noninteracting-blip approximation \cite{sn05,ns11,segal14}.
Numerically exact techniques, developed for the study of the (single-bath) SB model,
were similarly generalized to explore transport properties:
the multilayer multiconfiguration time-dependent Hartree theory
 \cite{kirilMCTDH}, influence
functional path integral techniques \cite{segal13} and Monte-Carlo simulations \cite{saitoMC}.


Theoretical studies of heat flow in model systems such as the NESB nanojunction  
are motivated by recent experiments of thermal energy flow across
alkane chains \cite{dlott1,dlott2,azulene}, proteins \cite{hamm1} and small aromatic molecules
\cite{rubtsov,dlott3,dlott4,dlott5}. These studies aim in exploring the role of vibrational energy flow in
e.g. chemical reaction dynamics, protein folding, conformational changes, and molecular electronics.
Questions of phononic heat transfer are of great interest in other disciplines.
For example, in thermoelectric applications reducing the phononic contribution to the thermal conductivity
improves the (heat to work) conversion efficiency;
recent experiments reached reduced thermal conductivities in nanocomposites \cite{thermoE}.
It is particularly interesting to design a molecular-level or a nanoscale thermal diode, optimally commanding
 unidirectional energy flow.
This would allow control over molecular reactivity, and potentially turn into a building block in
phononic (and similarly photonic)  devices.
Unidirectional heat flow was recently demonstrated in
nitrobenzene \cite{dlott3}: Using ultrafast infrared Raman spectroscopy it was shown that
 energy transfer from the nitro to the phenyl group,
or from the nitro to global vibrational modes, was blocked. However, vibrational energy was transferred
from the phenyl-localized modes to the nitro modes and to global modes.
It is now established that control over quantum energy flow can be achieved by combining many-body interactions
with spatial asymmetries \cite{sn05,lirev12}. Anharmonicity of vibrational modes is thus an essential ingredient for
building nontrivial functionalities. While more detailed calculations are imperative to explore particular
systems \cite{Leitner1,Leitner2}, the NESB model with a two-level system, a truncated harmonic vibration, 
is the simplest-nontrivial model which can allow us to explore the role of anharmonic (many-body) effects
in phononic  (or photonic) conduction.

In this work, we are interested in the
problem of quantum heat transfer in anharmonic nanojunctions, particularly
when the central object's coupling energy to the contacts is substantial.
Our goal is to examine and compare different techniques, understand their range
of validity, and find out when they provide qualitatively correct results in comparison
 to exact numerical techniques.
Focusing on the NESB model, we aim in rectifying the following points:

(i) Relation between dissipative dynamics and transport.
We study here both the RDM dynamics
and the transfer of heat in the NESB model using the weak-coupling (system-bath)
Bloch-Redfield (BR) formalism, the noninteracting-blip
approximation (NIBA), valid in the non-adiabatic limit and at high
temperatures, and  numerically-exact influence-functional path integral simulations. The BR and NIBA techniques provide consistent
results when describing the dynamics of the RDM
in a certain range of parameters.
Does this agreement translate into consistent transport properties?
The answer is negative. We show here that even when the BR and NIBA techniques reasonably agree
on the RDM dynamics,
results significantly deviate when following the heat current
behavior: The BR scheme fails in providing  the current
characteristics, qualitatively and quantitatively, beyond the very weak coupling limit.
Similarly, beyond the strict non-adiabatic limit NIBA badly fails in describing transport trends,
while it still performs reasonably well in RDM calculations.

(ii) Developing approaches for weak-intermediate coupling cases.
The BR method administers quantum kinetic equations, and it provides a transparent theory for thermal conduction:
a linear enhancement of current with increasing
coupling energy to the contacts. Other methods
reveal that this trend breaks down immediately beyond the very weak coupling limit 
\cite{ns11,vtw10,wu14,segal13,saitoMC}.
However, a careful comparison between different techniques is missing.
To study physical situations, e.g., with the molecule moderately or strongly attached to thermal contacts
 as in Ref. \cite{azulene},
 it is imperative to develop reliable methodologies that can extend beyond the very weak coupling regime.
We play here with four different approaches, BR \cite{sn05,claire09,wang12}, NIBA \cite{sn05,ns11},
perturbative techniques based on non-equilibrium Green's function (NEGF)
  \cite{vtw10,wu14},
and  numerically exact influence functional path-integral simulations \cite{segal13}.
We study the  current characteristics as a function of the contact interaction, as well as the
temperature and the frequency of the TLS, and observe a nontrivial non-monotonic performance
of the junction, exposing the underlying mechanisms of thermal conduction.


The paper is organized as follows. In Sec. \ref{model} we present
the model and observables of interest: the reduced
density matrix and the heat current, including the linear
response coefficient, the thermal conductance. In Sec. \ref{BR} we lay down the
Bloch-Redfield equations for the RDM and the steady-state
current. Sec. \ref{NIBA} presents the corresponding NIBA equations.
Sec. \ref{INFPI} describes influence functional path integral approaches.
Numerical results for the RDM dynamics and steady-state heat current are included in Sec. \ref{results}. In Sec.
\ref{Summ} we summarize our work.


\section{Model and Observables of Interest}
\label{model}

The NESB model includes a two-state system (spin) bridging two
bosonic reservoirs ($\nu=L,R$). In the ``local" basis ($|0\rangle$ and $|1\rangle$)
the isolated spin Hamiltonian reads
\bea
H_0={\hbar\omega_0\over 2} \sigma_z + {\hbar\Delta\over 2} \sigma_x,
\label{eq:H0}
\eea
and the total Hamiltonian is given by
\beq
H= H_0
+
\sum_{\nu,k} \left[
{\hbar\sigma_z\over 2}
\lambda_{k,\nu}
(b_{k,\nu}^{\dagger} + b_{k,\nu}) +\hbar \omega_{k} b_{k,\nu}^{\dagger} b_{k,\nu} \right].
\label{eq:ham}
\eeq
The Pauli matrices are defined as
$\sigma_z=|1\rangle\langle 1|-|0\rangle\langle 0|$,
$\sigma_x=|0\rangle\langle 1|+|1\rangle\langle 0|$, and
$\sigma_y=-i|1\rangle\langle 0|+i|0\rangle\langle 1|$,
$\hbar\omega_0$ is the level detuning (bias), $\Delta$ stands for
the tunneling frequency between the spin states, and
$b_{k,\nu}^{\dagger}$ ($b_{k,\nu}$) is the creation (annihilation)
operator of a boson (e.g. phonon) with a wavenumber $k$ in the
$\nu$ reservoir. The interaction of the subsystem with the
 baths can be characterized by a spectral density function,
defined as
\beq
J_{\nu} (\omega) = \sum_{k} \lambda_{k,\nu}^2 \, \delta (\omega - \omega_{k}).
\eeq
We perform our numerical simulations using an Ohmic form,
\beq
J_{\nu} (\omega) = 2\alpha_{\nu} \omega e^{-\omega/\omega_c}.
\label{eq:spect}
\eeq
The methodologies discussed below can handle other spectral functions.
Here $\alpha_\nu$ is a dimensionless interaction parameter
between the spin subsystem and the $\nu$ reservoir.
Below we use the definition $\alpha\equiv \alpha_L + \alpha_R$.
For simplicity, the cutoff frequency $\omega_c$ is taken identical
in both baths.
In the context of electron transfer processes it is useful to define
the reorganization energy $E_r^{\nu}\equiv \int d\omega J_{\nu}(\omega)/\omega$.
For Ohmic functions it reduces to  $E_r^{\nu}=2\alpha_{\nu} \omega_c$.

Below we principally work in the so-called non-adiabatic limit of $\Delta/\omega_c\ll 1$
and temperatures $k_BT/\hbar \Delta\gtrsim 1$.
The ``non-adiabatic" terminology is delivered from studies of electron transfer reactions in condensed phases,
in which the TLS represents electron-donor and acceptor states with a tunneling frequency $\Delta$:
``adiabatic processes" $\Delta>\omega_c$ refer to reactions with fast tunneling electrons 
relative to the phonon bath.
 In the opposite non-adiabatic limit $\Delta\ll\omega_c$ 
the characteristic time scale of the bath $1/\omega_c$ is short relative to the internal timescale for tunneling.  

Based on transport results in this range,
we identify four regions:
(i) we refer below to  $\alpha_{\nu}<0.025$  as the very weak coupling regime,
(ii) $0.025<\alpha_{\nu}<0.1$ corresponds to the weak coupling limit,
(iii) $0.1<\alpha_{\nu}<0.5$ describes the intermediate regime, and (iv) $\alpha_{\nu}>0.5$
corresponds to the strong coupling limit.


The reduced density matrix is organized from the population difference
$\langle \sigma_z(t)\rangle$ and the real and imaginary parts of the coherence $\langle \sigma_{x,y}(t)\rangle$,
\bea \langle \sigma_i(t)\rangle = {\rm tr}\{
\rho(t=0)e^{iHt/\hbar}\sigma_ie^{-iHt/\hbar} \},
\eea
with $\rho(t=0)$ as the initial state of the total density matrix.
In what follows we assume a factorized initial condition,
$\rho(t=0)=\rho_L\otimes\rho_R\otimes|1\rangle\langle 1|$. 
The two reservoirs, $H_{\nu}=\sum_{k}  \hbar \omega_{k} b_{k,\nu}^{\dagger}
b_{k,\nu}$, are separately prepared in a canonical-equilibrium state
of temperature $T_{\nu}=1/(k_B\beta_{\nu})$,
\bea \rho_{\nu}=\frac{e^{-\beta_{\nu} H_{\nu}}}{Z_{\nu}}, \,\,\,\,
Z_{\nu}={\rm Tr}_{\nu}e^{-\beta_{\nu}H_{\nu}}.
\eea
At $t=0$ the subsystem-bath interaction is turned on and
we wait for the (assuming unique) steady-state solution to set in. The heat current can be computed
from the transient regime  to the steady-state limit; below we
focus only on the long-time behavior. It is reached
by considering energy leakage at the contacts.
For example, at the left contact the heat current operator is defined as
\bea \hat{j}_L \equiv \frac{dH_L}{dt}= \frac{i}{\hbar}[H,H_L].
\label{eq:defjq}
\eea
Tracing over all degrees of freedom we reach the expectation value
\bea
j_L\equiv{\rm tr}[\rho(t=0)\hat j_L(t)].
\label{eq:defjq2}
\eea
Operators are written here in the Heisenberg representation. In
steady-state, $j_q=j_L=-j_R$. In the linear response regime the
current is expanded to the lowest order in the temperature difference,
$j_q\sim\kappa(T_L-T_R)$, and we obtain the thermal
conductance $\kappa$ from the relation
\bea
\kappa \equiv \frac{ d j_q}{ d T_L} \Big|_{ T_L\to T_R=T}.
\label{eq:kappadef}
\eea
To practically compute  the heat current
we wish to manipulate Eq. (\ref{eq:defjq2}) into a workable definition.
A formally-exact construction 
has been derived in Ref. \cite{vtw10} from the
perturbation expansion of the non-equilibrium Green's function. This
formula, a many-body extension of  Landauer's expression \cite{rk98},
an analog of the Meir-Wingreen formula for electronic systems \cite{mw92}, expresses
the heat current of the NESB model in correlation functions of the
spin.
The linear response limit of this formula was recently studied using Monte-Carlo
simulations to explore signatures of Kondo physics in thermal conduction \cite{saitoMC}.
In the following sections we describe the evaluation of the  heat current within different sets of approximations.

%

\section{Bloch-Redfield Formalism: Very weak coupling regime}
\label{BR}

\subsection{Hamiltonian}
\label{BRHam}

The standard Bloch-Redfield equation can be derived from the exact quantum master equation based on the assumption
of weak system-bath interactions. It is convenient to develop it
 in the ``energy" basis, the representation in which
the spin subsystem is diagonal,
\bea
H_S=\sum_n E_{n}|n\rangle \langle n|.
\label{eq:BRHS}
\eea
%
In addition, the total Hamiltonian should be prepared in an additive structure,
\bea
\tilde H=H_S+H_L+H_R + V_L+V_R,
\label{eq:HBR}
\eea
with the thermal baths denoted each by $H_{\nu}$ and the system-bath
interaction given in a direct-product form,
\bea
V_{\nu}&=&S^{\nu}\otimes B_{\nu},
\nonumber\\
S^{\nu}&=&\sum_{n,n'} S^{\nu}_{n,n'}|n\rangle \langle n'|.
\label{eq:V}
\eea
Here $B_{\nu}$ and $S^{\nu}$ are bath and  subsystem operators,
respectively. 
We label the subsystem operator by the index $\nu=L,R$;  the impurity may couple to the two baths via distinct
operators.

Specifically to the NESB model, we diagonalize the isolated TLS of
Eq. (\ref{eq:H0}) with a rotation matrix
$U=e^{-\frac{i}{2}\theta \sigma_y}$, $\tan \theta=\Delta/\omega_0$.
The total Hamiltonian (\ref{eq:ham}) transforms into $\tilde H\equiv U^{\dagger}
HU$, with
\bea
\tilde H&=&
{\hbar\epsilon\over 2} \tilde \sigma_z +
\sum_{\nu,k}\hbar \omega_{k} b_{k,\nu}^{\dagger} b_{k,\nu}
\nonumber\\
&+& \frac{1}{2}\left( \hbar\tilde\sigma_z \cos\theta
-\hbar\tilde\sigma_x\sin\theta \right) \sum_{\nu,k}
\lambda_{k,\nu}(b_{k,\nu}^{\dagger} + b_{k,\nu}), \label{eq:hamBR}
\eea
where $\epsilon=\sqrt{\omega_0^2+\Delta^2}$ and $\tilde\sigma_i$ are the Pauli matrices
in the energy basis, denoted here by $|\pm\rangle$. We now identify the
operators of Eq. (\ref{eq:V}) by
\bea
S^{\nu}&=& 
\tilde\sigma_z \cos\theta -\tilde\sigma_x\sin\theta, 
\nonumber\\
B_{\nu}&=&\frac{1}{2}\sum_{k} \hbar\lambda_{k,\nu}(b_{k,\nu}^{\dagger} + b_{k,\nu}) .
\label{eq:SB}
\eea
The dynamics of the reduced density matrix under the BR equation
has been examined in numerous studies \cite{legget, Aslangul}. For
completeness, we include in Sec. \ref{BRD}  relevant expressions using
a notation similar to that employed in Ref. \cite{thoss01}.
The heat current in the
NESB model was only recently derived in a closed form  at the level of the BR scheme
\cite{claire09,wang12}. A workable expression is provided in Sec. \ref{BRH}.

\subsection{Reduced density matrix}
\label{BRD}

In the BR scheme the interaction $V_{\nu}$ 
is treated perturbatively, to the lowest nontrivial order \cite{ARedfield}. This results in
a master equation for the spin RDM \cite{silbey80}.
It obeys an integro-differential equation,
written here in the local-site basis of Eq.
(\ref{eq:ham}) \cite{silbey80,Aslangul,thoss01},
\bea
\frac{d}{dt}\langle\sigma_z(t)\rangle&=&\Delta\langle\sigma_y(t)\rangle,
\nonumber\\
\frac{d}{dt}\langle\sigma_y(t)\rangle&=&
\omega_0\langle\sigma_x(t)\rangle-\Delta\langle\sigma_z(t)\rangle-
\int_0^t d\tau G_y(\tau)
-\int_0^t d\tau\left[G_{yy}(\tau)\langle\sigma_y(t-\tau)\rangle+G_{yx}(\tau)\langle\sigma_x(t-\tau)\rangle\right],
\nonumber\\
\frac{d}{dt}\langle\sigma_x(t)\rangle&=
&-\omega_0\langle\sigma_y(t)\rangle-\int_0^t d\tau G_x(\tau)-
\int_0^t d\tau\left[G_{xx}(\tau)\langle\sigma_x(t-\tau)\rangle
-G_{yx}(\tau)\langle\sigma_y(t-\tau)\rangle\right].
\label{eq:BRdyn}
\eea
The kernels satisfy
\bea
G_x(t) &=& \frac{\Delta}{\epsilon}\sin \left(\epsilon t\right)\sum_{\nu}M_{\nu}''(t), \,\,\,\
G_y(t) = \frac{\omega_0\Delta}{\epsilon^2}\left[1-\cos\left(\epsilon t\right)\right]\sum_{\nu}M_{\nu}''(t)
\nonumber\\
G_{xx}(t) &=& \cos\left(\epsilon t\right)\sum_{\nu}M_{\nu}'(t), \,\,\,\,\
G_{yy}(t) = \frac{\Delta^2+\omega_0^2\cos \left(\epsilon t\right)}{\epsilon^2} \sum_{\nu}M_{\nu}'(t)
\nonumber\\
G_{yx}(t) &=& \frac{\omega_0}{\epsilon}\sin\left(\epsilon t\right) \sum_{\nu}M_{\nu}'(t),
\eea
and the dissipative terms enclose the correlation function $\langle B_{\nu}(t)
B_{\nu}(0)\rangle_{\nu}=M_{\nu}'(t)-iM''_{\nu}(t)$, with
\bea
M_{\nu}'(t) &=& \int_0^\infty d\omega J_{\nu}(\omega)\coth(\beta_{\nu}\hbar\omega/2)\cos(\omega t),
\nonumber\\
M_{\nu}''(t) &=& \int_0^\infty d\omega J_{\nu}(\omega)\sin(\omega t).
\eea
%
Under the Markov approximation, we write a
time-local Markovian equation
$d\langle \sigma_i\rangle/dt=\sum_{i,j}D_{i,j}\sigma_{j}$, see e.g. \cite{thoss01,silbey80,RatnerCD},
and solve it in the long time limit.
The equilibrium ({\it eq}) solution, in the case of a single bath, is
 $\langle \sigma_z\rangle_{eq}=-\frac{\omega_0}{\epsilon}\tanh(\beta\hbar\epsilon/2)$,
  $\langle \sigma_x\rangle_{eq}=-\frac{\Delta}{\epsilon}\tanh(\beta\hbar\epsilon/2)$,
 and
$\langle \sigma_y\rangle_{eq}=0$.
In the non-equilibrium two-bath scenario the steady-state ({\it ss}) solution of
the Markovian equation (reached at $t\rightarrow \infty$) satisfies
\bea
\langle \sigma_z\rangle_{ss}&=&
-\frac{\omega_0}{\epsilon}\frac{J_L(\epsilon)+ J_R(\epsilon)}{J_L(\epsilon)\coth(\beta_{L}\hbar\epsilon/2)+ J_R(\epsilon)
\coth(\beta_R\hbar\epsilon/2)},
\nonumber\\
\langle \sigma_x\rangle_{ss}&=&\frac{\Delta}{\omega_0}\langle \sigma_z\rangle_{ss},
\nonumber\\
\langle \sigma_y\rangle_{ss}&=& 0,
\eea
reducing to the correct equilibrium state \cite{legget}.
These elements build-up the reduced density matrix,
\bea
\rho_S^{ss}
=\frac{1}{2} 
\left( \begin{array}{cc}
1+\langle \sigma_z\rangle_{ss} &  \langle \sigma_x\rangle_{ss}-i \langle \sigma_y\rangle_{ss}  \\
\langle \sigma_x\rangle_{ss}+i\langle \sigma_y\rangle_{ss} & 1-\langle \sigma_z\rangle_{ss}  \\
\end{array} \right). 
\label{eq:BRss}
\eea
We now introduce a short notation,
$x\equiv\frac{J_L(\epsilon)+ J_R(\epsilon)}{J_L(\epsilon)\coth(\beta_{L}\hbar\epsilon/2)+ J_R(\epsilon)
\coth(\beta_R\hbar\epsilon/2)}$, and condense the RDM into
\bea \rho_S^{ss}=\frac{1}{2}\hat I
-\frac{x}{\epsilon}\left(\frac{\omega_0}{2}\sigma_z  +
\frac{\Delta}{2}\sigma_x \right).
\label{eq:BRss2}
\eea
Here $\hat I$ stands for the
identity matrix. In the energy basis $|\pm\rangle$, $\tilde\rho_S^{ss} =
U^{\dagger} \rho_{S}^{ss}U$, resulting in the simple form
\bea
\tilde\rho_S^{ss} = \frac{1}{2}\hat I -\frac{x}{2}\tilde\sigma_z.
\label{eq:BRRDMtilde}
\eea
Explicitly, the population of the two levels in the energy basis follows
\bea \tilde p_-^{ss}&=& \frac{J_L(\epsilon)[1+n_L(\epsilon)]+
J_R(\epsilon)[1+n_R(\epsilon)]} {J_L(\epsilon)[1+2n_L(\epsilon)]+
J_R(\epsilon)[1+2n_R(\epsilon)]},
\nonumber\\
\tilde p_+^{ss}&=&1-\tilde p_-^{ss},
\label{eq:BRpop}
\eea
with the Bose-Einstein function $n_{\nu}(\epsilon)=[e^{\beta_{\nu}\hbar\epsilon}-1]^{-1}$.
At thermal equilibrium, $T_L=T_R$, the spin occupation
depends on the temperature of the bath and
the spin splitting, but it does carry information on the coupling strength of the spin to the bath.
In contrast, it is significant to observe signatures
of the non-equilibrium situation $T_L\neq T_R$
in the  levels' population, now controlled by the spectral functions of the baths.
Thus, out-of-equilibrium the coupling energy to the contacts  not only determines the relaxation rate towards steady-state,
but it further dictates the steady-state solution.
The functional form (\ref{eq:BRpop}) has been obtained in Refs.
\cite{sn05,ns11,saitoMC} considering the {\it unbiased} spin-boson model,
$\omega_0=0$. Thus, the seemingly
more complex-biased model does not expose a nontrivial $\omega_0$-dependent controllability.

Equations (\ref{eq:BRdyn}) and (\ref{eq:BRss2}) complete our
discussion of the RDM under the Bloch-Redfield formalism:
dynamics and steady-state solution.
In the next subsection we use these expressions and calculate the steady-state heat current  in the NESB model.

\subsection{Heat Current}
\label{BRH}

A closed expression for  quantum heat conduction in multi-state nanojunctions
has been derived in Ref. \cite{claire09} under a second order perturbation expansion
in the system-bath coupling and the rotating wave approximation.
This result was recently extended in Ref. \cite{wang12} to include anti-rotating wave terms, transient
effects, and lamb shifts of energies.
This derivation is not repeated here; we only review its principles.
Briefly, the expectation value of the heat current operator is
attained from the definition (\ref{eq:defjq}) by time-evolving the
density matrix in the energy basis to {\it first} order in the system-bath
interaction term $V_{\nu}$. The overall result is second-order in the interaction
parameter since the definition of the heat current itself includes
the system-bath interaction operator.
We trace the current operator over the baths with a factorized system-bath initial
density matrix and take the long time limit. The resulting Bloch-Redfield-type steady-state heat
current formula is given by the simple form \cite{wang12}
%
%
%
%
\bea j_{\nu}={\rm Tr_{S}}[\tilde \rho_S^{ss}A^{\nu}].
\label{eq:BRCurrent}
\eea
The relevant RDM is provided in Eq. (\ref{eq:BRRDMtilde}).
The matrix $A^{\nu}$
depends on the properties of the subsystem;  the $\nu$
index marks the terminal in which the current is calculated,
\bea
A^{\nu}_{k,j}=\sum_{l}S^{\nu}_{k,l}
S^{\nu}_{l,j} \left( E_{l,j}w_{j\to l}^{\nu} + E_{l,k}(w_{k \to l}^{\nu})^*\right).
\label{eq:BRA}
\eea
Here $E_{k,l}=E_k-E_l$,  with the subsystem eigenenergies Eq. (\ref{eq:BRHS}).
The rate constants are given by half-range Fourier
transforms of bath correlation functions,
\bea
w_{j\to l}^{\nu}&=&\frac{1}{\hbar^2}
\int_0^{\infty}d\tau e^{-iE_{l,j}\tau/\hbar}
\langle B_{\nu}(\tau)B(0)\rangle_{\nu}
\nonumber\\
&=& \frac{1}{4}\int_0^{\infty}d\tau e^{-iE_{l,j}\tau/\hbar}
\int_0^{\infty} d\omega J_{\nu}(\omega)\left[ e^{i\omega\tau
}n_{\nu}(\omega)+ e^{-i\omega\tau }(n_{\nu}(\omega)+1)\right],
 \eea
where the real part satisfies
\bea
\Re[w_{j\to l}^{\nu}]= \frac{\pi}{4}J_{\nu}(|E_{l,j}|/\hbar) \times
\begin{cases}
n_{\nu}(E_{l,j}/\hbar) & {\rm if} \, E_l>E_j\\
[n_{\nu}(E_{j,l}/\hbar)+1] & {\rm if} \,E_l<E_j.
\end{cases}
\eea
Since in the energy basis the reduced density matrix of the NESB model is diagonal
in steady-state, see Eq. (\ref{eq:BRRDMtilde}),
the current in Eqs. (\ref{eq:BRCurrent})-(\ref{eq:BRA}) immediately simplifies to
\bea
j_{\nu}=2 \sum_{l,k} |S_{k,l}^{\nu}|^2E_{l,k} (\tilde \rho_S^{ss})_{k,k}\Re[w_{k\to l}^{\nu}].
\eea
Explicitly, for the two-state system this brings out an intuitive structure,
\bea
j_{\nu}&=&
2|S_{-,+}^{\nu}|^2
E_{+,-} \left\{ \tilde p_-^{ss}\Re[w_{-\to +}^{\nu}]
-  \tilde p_+^{ss}\Re[w_{+\to -}^{\nu}] \right\}
\nonumber\\
&=& (\hbar \epsilon) \sin^2(\theta)   \Gamma_{\nu} (\epsilon) \left\{ \tilde p_-^{ss} n_{\nu}(\epsilon)
- \tilde p_+^{ss} [n_{\nu}(\epsilon)+1] \right\}.
\label{eq:jBR}
\eea
It describes the current, say at the $L$ contact, by the net process of an $L$-bath induced excitation, multiplied by
the ground state population and the energy difference $E_{+,-}$, and the relaxation from the excited state,
to dispose energy in the $L$ bath.
The second line was reached from the definition $\Gamma_{\nu}(\epsilon)\equiv\frac{\pi}{2} J_{\nu}(\epsilon)$,
and by identifying the energy difference $E_{+,-}=\hbar \epsilon$.
Employing the steady-state occupations (\ref{eq:BRpop}) we obtain the closed-form result
\bea j_q=(\hbar\epsilon)\sin^2(\theta)
\frac{
\Gamma_L(\epsilon)\Gamma_R(\epsilon)}
{\Gamma_L(\epsilon)[1+2n_L(\epsilon)]+\Gamma_R(\epsilon)[1+2n_R(\epsilon)]} [n_L(\epsilon)-n_R(\epsilon)].
\label{eq:jweak} \eea
%
%
The heat current can be expanded in orders of $\Delta T=T_L-T_R$ to yield
the thermal conductance, a linear response coefficient (\ref{eq:kappadef}),
\beq \kappa &=& \sin^2\theta \frac{(\hbar\epsilon)^2}{k_BT^2}
\frac{\Gamma_R(\epsilon)
\Gamma_L(\epsilon)}{[\Gamma_R(\epsilon)+\Gamma_L(\epsilon)][1+2n(\epsilon)]}\frac{e^{\hbar\epsilon\beta}}
{(e^{\hbar\epsilon\beta}-1)^2}.
\eea
%
%
Here $n(\omega)=[e^{\beta\hbar\omega}-1]^{-1}$ denotes the
Bose-Einstein distribution function at the inverse temperature
$\beta=1/(k_BT)$. Note that $\Delta^2/\epsilon^2=\sin^2\theta$.
So far our discussion did not assume a particular spectral function. We now employ
the Ohmic form (\ref{eq:spect})
with a large cutoff (non-adiabatic limit), $\omega_c\gg \epsilon$, and receive \cite{saitoMC}
\beq
\kappa &=& \pi\epsilon\frac{(\hbar \Delta)^2}{k_BT^2}
{ \alpha_L\alpha_R  \over (\alpha_L+\alpha_R) }
\frac{ 1}{2\sinh(\hbar\epsilon\beta)}
\nonumber\\
&\xrightarrow{\beta\hbar\epsilon\ll1}&
\frac{\hbar\pi}{2} \frac{\Delta^2}{T}
\frac{\alpha_L\alpha_R}{\alpha_L+\alpha_R}.
\label{eq:BRkappa2}
\eeq
In the classical high temperature limit the
thermal conductance is identical for  biased ($\omega_0\neq0$)
and unbiased ($\omega_0=0$) models. Furthermore, the prefactor $\Delta^2$ evinces
on the expected agreement of the BR expression with the NIBA formalism for Ohmic baths,
the latter technique strictly holds only in the non-adiabatic regime ($\Delta\ll\omega_c$).




\section{Noninteracting-Blip Approximation: Strong Coupling}
\label{NIBA}

NIBA equations were derived for describing
spin polarization dynamics in the SB model based on a path integral influence
functional formalism \cite{legget,weiss}. Alternatively, these equations can be
 recovered by transforming the SB Hamiltonian to the shifted-polaron representation, then
working out a second order perturbation theory expansion of the RDM in the dressed tunneling
frequency \cite{Aslangul,dekker}.
NIBA serves as a good approximation for the spin polarization in the non-adiabatic limit
$\omega_c>>\Delta$ \cite{weiss}. For Ohmic baths, it is exact for $\langle \sigma_z(t)\rangle$
for the unbiased model at weak damping, and it can
faithfully simulate the SB dynamics (polarization and coherences) at strong system-bath interactions and/or at high
temperatures.

In a series of recent studies we had extended NIBA
to the out-of-equilibrium regime
\cite{sn05,segal06,ns11,segal14} aiming in simulating
heat transport in the NESB nanojunction beyond the BR  weak coupling limit.
This was achieved by writing down
the cumulant generating function of the system, to derive
a NIBA-formula for the heat current \cite{sn05,ns11}.
We recently proved that the approximate NIBA expression
agrees with numerically-exact simulations of the thermal conductance \cite{saitoMC} for
Ohmic reservoirs, working in the high temperature limit.

In Sec. \ref{NIBAR} we include equations of motion for the spin RDM under NIBA \cite{weiss}.
The NIBA heat current expression was first constructed in Ref. \cite{sn05}; it was later
formally derived from a counting-statistics approach in Ref. \cite{ns11}. In Sec. \ref{NIBAC}
we include relevant expressions.

\subsection{Reduced density matrix}
\label{NIBAR}

The exact-formal series expansion
of $\langle \sigma_z(t)\rangle$ in $\Delta$ can be organized as a generalized master equation.
The dynamics of off-diagonal terms $\langle \sigma_{x,y}(t)\rangle$
is obtained from the polarization $\langle \sigma_z(t)\rangle$
by exact integral relations \cite{weiss},
\bea
\frac{d \langle \sigma_z\rangle}{dt}&=& -\int_{0}^{t}
K_{s,z}(t-\tau)\langle \sigma_z(\tau)\rangle d\tau -\int_{0}^{t}
K_{a,z}(t-\tau)d\tau,
\nonumber\\
\langle \sigma_x\rangle&=& \int_{0}^{t}
K_{s,x}(t-\tau) d\tau -\int_{0}^{t}
K_{a,x}(t-\tau)\langle\sigma_z(\tau)\rangle d\tau,
\nonumber\\
\langle \sigma_y\rangle&=&\frac{1}{\Delta} \frac{d \langle \sigma_z\rangle}{dt}.
\label{eq:NIBAintegro}
\eea
The (exact) kernels in Eq. (\ref{eq:NIBAintegro}) can be truncated to include lowest-nontrivial terms in $\Delta$.
This scheme, termed the noninteracting-blip approximation, results in
 \cite{weiss},
\bea K_{s,z}(t)&=&\Delta^2e^{-Q'(t)}\cos[Q''(t)]\cos(\omega_0 t),
\nonumber\\
K_{a,z}(t)&=&\Delta^2e^{-Q'(t)}\sin[Q''(t)]\sin(\omega_0 t),
\nonumber\\
K_{s,x}(t)&=&\Delta e^{-Q'(t)}\cos[Q''(t)]\sin(\omega_0 t),
\nonumber\\
K_{a,x}(t)&=&\Delta e^{-Q'(t)}\sin[Q''(t)]\cos(\omega_0 t).
\label{eq:NIBAK}
\eea
The function $Q(t)=\sum_{\nu}Q_{\nu}(t)$,
$Q_{\nu}(t)=Q_{\nu}'(t)+iQ_{\nu}''(t)$ contains real and imaginary components with
\bea
Q'_{\nu}(t)& = & \int_{0}^{\infty}d\omega\frac{J_{\nu}(\omega)}{\omega^2}[1-\cos(\omega t)] [1+2n_{\nu}(\omega)],
\nonumber\\
Q''_{\nu}(t)& = &  \int_{0}^{\infty}d\omega \frac{J_{\nu}(\omega)}{\omega^2}\sin(\omega t).
\label{eq:NIBAQ}
\eea
In an equilibrium  situation, $T_L=T_R$, $\langle
\sigma_x\rangle_{eq}=-\frac{\Delta}{\omega_0}\tanh\left(\frac{\beta\hbar\omega_0}{2}\right)$ and
$\langle \sigma_z\rangle_{eq}=-\tanh\left(\frac{\beta\hbar\omega_0}{2}\right)$
\cite{legget}. The out-of-equilibrium steady-state solution of the Markovian equation satisfies
\cite{sn05,ns11}
\bea
\langle \sigma_z\rangle_{ss}=[k(-\omega_0) - k(\omega_0) ]/ [k(-\omega_0) + k(\omega_0)],
\eea
where the rates are convolutions of the $L$ and $R$ baths-induced rates,
\bea
k(\omega_0)&=&\int_{-\infty}^{\infty}e^{i\omega_0 t} e^{-Q_{L}(t)}e^{-Q_{R}(t)}dt
\nonumber\\
&=&\frac{1}{2\pi}\int_{-\infty}^{\infty}k_L(\omega_0-\omega)k_R(\omega)d\omega.
\label{eq:NIBAconv}
\eea
The ``Fermi-Golden-Rule" rate constant 
$k_{\nu}(\omega)$  ($\times \Delta^2$) was originally derived in the context 
of reaction rates in donor-acceptor complexes \cite{nitzan},
\bea
k_{\nu}(\omega)=\int_{-\infty}^{\infty}e^{i\omega t} e^{-Q_{\nu}(t)}dt,
\label{eq:rate}
\eea
and it satisfies the detailed balance relation,
\bea
k_{\nu}(-\omega)=k_{\nu}(\omega)e^{-\beta\hbar\omega}.
\label{eq:ratedb}
\eea
In the Ohmic case, in the scaling regime,
$k_BT,\hbar\omega < \hbar\omega_c$, the $\nu$-bath-induced rate obeys
\cite{weiss}
\bea
k_{\nu}(\omega)=
\frac{1}{\omega_c} \left( \frac{\hbar \omega_c}{2\pi k_BT} \right)^{1-2\alpha_{\nu}}
\frac{|\Gamma(\alpha_{\nu}+i\hbar \omega / 2\pi k_BT)|^2}
{\Gamma(2\alpha_{\nu})}e^{\hbar\omega/2k_BT}.
\label{eq:NIBAk1}
\eea
%
Closed expressions for $k(\omega)$, thus the polarization, are missing in general since the convolution
(\ref{eq:NIBAconv}) is nontrivial to handle analytically.
As we show below, NIBA heat current relies on the
steady-state population of the spin states (in the local basis),
$p_{1}^{ss}=(1+ \langle \sigma_z\rangle_{ss})/2$,  $p_0=1-p_1$,
but it does not contain the coherences. In equilibrium,
%
the spin occupation obeys
\bea
p_1^{eq}
= \frac{e^{-\beta\hbar\omega_0/2}}{e^{-\beta\hbar\omega_0/2} + e^{\beta\hbar\omega_0/2}}.
\label{eq:NIBApop}
\eea
%
It is of interest to acquire $p_{1,0}^{ss}$ approximately-analytically, to study signatures of the non-equilibrium
condition on the TLS at strong system-bath couplings.

\subsection{Heat current}
\label{NIBAC}

A closed expression for the steady-state heat current under NIBA
has been derived by energy-unraveling the polarization dynamics, 
Eq. (\ref{eq:NIBAintegro}) \cite{ns11}, resulting in
%
\bea j_q = \left(\frac{\Delta}{2}\right)^2
\frac{\hbar}{2\pi}\int_{-\infty}^{\infty}\omega d\omega \left[
k_R(\omega)k_L(\omega_0-\omega)p_1^{ss} -
k_R(-\omega)k_L(-\omega_0+\omega)p_0^{ss} \right].
\label{eq:NIBACurrent} \eea
Considering the unbiased model, the thermal conductance is given by the compact form \cite{segal14}
\bea
\kappa
=
\left(\frac{\Delta}{2}\right)^2 \frac{\hbar^2}{4\pi k_B T^2}
\int_{-\infty}^{\infty}\omega^2
 k_R(\omega)k_L(-\omega)d\omega.
\label{eq:NIBAkappaw0}
\eea
%
In the high $T$ limit we substitute Eq. (\ref{eq:NIBAk1}) into Eq. (\ref{eq:NIBAkappaw0}) and receive \cite{segal14}
\bea
\kappa &\simeq& \mathcal A \hbar^2\left(\frac{\Delta}{\omega_c}\right)^2
\frac{1}{k_BT^2}
\left(\frac{\hbar \omega_c}{ k_B T}\right)^{2-2\alpha_L-2\alpha_R}
\int_0^{k_BT/\hbar} \omega^2d\omega
\nonumber\\
&\simeq& \mathcal A
\frac{k_B\Delta^2}{\omega_c}
\left(\frac{\hbar \omega_c}{ k_B T}\right)^{1-2\alpha}.
\label{eq:kapNiba}
\eea
%
We emphasize that under NIBA the heat current, Eq.
(\ref{eq:NIBACurrent}), is only determined by the polarization dynamics and its decay rates to steady-state.
This observation explains the satisfactory
performance of NIBA (with $\alpha$) in heat transfer calculations \cite{segal14}:
NIBA truncation of $K_{s/a,z}$
carries errors (in the inter-blip correlations of the kernel) only second-order in
$\alpha$. As a result, the dynamics of $\langle\sigma_z(t)\rangle$
is exact for the unbiased and weakly-damped NESB model. In contrast,
the kernels $K_{s/a,x}$ carry first-order errors in $\alpha$, making them inaccurate even for
the unbiased SB model. However, these errors do not propagate into the NIBA heat current formula.
%


\section{Path Integral Simulations: QUAPI and INFPI}
\label{INFPI}

The numerically exact iterative quasi-adiabatic path-integral (QUAPI) approach
has been developed by Makri and Makarov for simulating the reduced spin dynamics in
the spin-boson model \cite{QUAPI}. It allows one
to treat strong system-bath couplings and to include non-Markovian effects.
Recent efforts directed excitonic energy transfer in biomolecules, treating more complex
situations,  considering (a single excitation in) multiple sites and prominent
vibrations in each site \cite{thorwartE}.

The QUAPI algorithm had been constructed on the grounds
of harmonic environments linearly coupled to the subsystem \cite{QUAPI}.
In this situation,
the effect of the environment on the subsystem's RDM
can be absorbed in an analytic function, the ``Feynman-Vernon influence functional" \cite{FV}.
At nonzero temperatures
the memory function within the influence functional
decays rapidly in time,
allowing for its controlled truncation and the development of an iterative
time evolution scheme \cite{QUAPI}.

This principle can be generalized to construct a QUAPI-type algorithm
for  anharmonic baths \cite{Makri-An}.
Furthermore, a related-general approach, the so-called ``influence functional path
integral" (INFPI) tool has been put forward for treating quantum systems in
contact with multiple fermionic reservoirs at finite bias voltages \cite{INFPI1}.
This approach can be excersized when an exact analytic form for the influence functional is missing, as it is computed  numerically using trace identities.
Given their conceptual similarity, QUAPI and INFPI were recently combined
for a unified study of charge transport
and vibrational excitation and dissipation in donor-acceptor molecular electronic diodes \cite{INFPI2}.

The standard QUAPI algorithm administers only the dynamics of the reduced density matrix \cite{QUAPI}.
The calculation of other observables, particularly the heat current in the NESB model,
necessitates significant technical advances.
In contrast, the INFPI algorithm has been formulated for treating a generic impurity Hamiltonian 
\cite{INFPI1,INFPI2} and
 with little effort it can be employed for the study of other observables of quadratic structure, beyond
the RDM. For example, INFPI has been applied for the investigation of charge current \cite{INFPI1} and equilibration dynamics
\cite{manas} in the single-impurity Anderson model,
More recently, we adopted INFPI to simulate qubit-mediated energy flow between metals \cite{segal13}.

Algorithmic details of QUAPI \cite{QUAPI} and INFPI \cite{INFPI1,INFPI2} can be found elsewhere.
Here  we only highlight their working principles and associated numerical errors.
The starting point of these techniques involves the Trotter factorization of
the time evolution operator, e.g.,
 into a free subsystem term and bath-related propagators.
After collecting the environmental contributions and tracing over the baths we
reach an influence functional-type expression, see Eq. (\ref{eq:dynamics}) below.
As mentioned above, at finite temperatures and/or a nonzero chemical potential bias
 bath correlations exponentially decay in time \cite{Mitraspin,ISPI,SMarcus},
allowing for their truncation beyond a memory time $\tau_c$.
An iterative time evolution scheme can then be constructed
by defining an auxiliary quantity (an extension of the observable of interest) on the time-window $\tau_c$.
This time-nonlocal object can be iteratively evolved from the initial condition to the final time $t$.

Simulations with QUAPI and INFPI involve two numerical errors:
(i) Trotter error due to the finite time-step adopted in the Trotter breakup $\delta t$,
and (ii) an error associated with the truncation of the influence functional to
cover a certain time window $\tau_c$.
Furthermore, in INFPI the Fermi sea is discretized.
Thus, one should confirm that within the relevant simulation time
the number of bath states 
does not affect results.


\subsection{Reduced density matrix}
\label{INFPIR}

The original QUAPI algorithm provided the subsystem's RDM when coupled to a single harmonic bath.
Here QUAPI is (trivially) extended to accommodate two reservoirs of different temperatures.
The time evolution of the RDM up to $t=N\delta t$, $N$ is an integer,
can be represented in a path integral formulation as \cite{QUAPI},
\bea
&&\langle s_N^+|\rho_S(t)|s_N^-\rangle=
\int ds_0^{\pm} \int ds_1^{\pm} ... \int ds_{N-1}^{\pm}
 I^{har}(s_0^{\pm}, s_1^{\pm},...,s_{N}^{\pm}).
\label{eq:dynamics}
\eea
Here $s_k^{\pm}$ represent the discrete path of the subsystem on the forward ($+$)
and backward ($-$) contours (not to be confused with the eigenstates
of $H_S$ described in Sec. \ref{BRHam}). As an initial condition we assume that
$\rho(0)=\rho_{L}\otimes\rho_R\otimes \rho_S(0)$ with the baths separated from the subsystem.
The integrand in Eq. (\ref{eq:dynamics}) is refereed to as an ``Influence Functional" (IF) \cite{FV}; note that
in Refs. \cite{FV,QUAPI} it was identified without the free-subsystem evolution terms.
For a harmonic bath bilinearly coupled to the subsystem
the IF is given by an exponential of a quadratic structure, multiplied by free subsystem propagation terms,
\bea
I^{har}(s_0^{\pm},...,s_{N}^{\pm})&=&\exp\Big[-\sum_{\nu}\sum_{k}^N \sum_{k'=0}^k(s_k^+-s_k^-)(\eta_{k,k'}^{\nu}s_{k'}^+ -
\eta_{k,k'}^{\nu *}s_{k'}^- ) \Big]
\nonumber\\
&\times&
\langle s_N^+|e^{-iH_0\delta t} |s_{N-1}^+\rangle ... \langle s_0^+ |\rho_S(0)|s_0^-\rangle ...
\langle s_{N-1}^-| e^{iH_0\delta t} | s_N^-\rangle.
\label{eq:harmonic}
\eea
The free Hamiltonian $H_0$ is defined in Eq. (\ref{eq:H0}),
the coefficients $\eta_{k,k'}^{\nu}$ depend on the spectral function of the $\nu$ bath
and its temperature. They were derived in Ref. \cite{QUAPI} by discretizing the Feynman-Vernon IF.

%

\subsection{Heat Current}
\label{INFPIT}

The heat current in the NESB model can be written as a Meir-Wingreen
formula combining spin-spin correlation functions \cite{vtw10}.
Here, rather than generalizing QUAPI to follow
correlation functions in a non-equilibrium setup, a nontrivial task,
we study the behavior of the NESB model through the related fermionic model, naturally
handled by INFPI.

The spin-boson Hamiltonian can be reached by bosonizing the spin-fermion (SF)
model, comprising a spin [Eq. (\ref{eq:H0})] and two metallic leads $\nu=L,R$, 1D electron gases
with linear dispersion. The metals are
prepared at different temperatures but at the same chemical potential. They
are connected indirectly, only through the TLS, blocking charge transfer between the metals but allowing for
energy flow.
The total non-equilibrium spin-fermion (NESF) Hamiltonian reads
\bea
H_{SF}= 
H_0+ \hbar\sigma_z
\sum_{\nu,p,p'}g_{p,\nu; p',\nu}  c_{p,\nu}^{\dagger}c_{p',\nu}
+ \sum_{\nu,p}\epsilon_{p,\nu} c_{p,\nu}^{\dagger}c_{p,\nu}.
\label{eq:SF}
\eea
Here $c_{p,\nu}^{\dagger}$ ($c_{p,\nu}$) represents a fermionic creation (annihilation) operator.
The spin polarization couples to intra-bath electron-hole pair generation
as in Ref. \cite{segal13}.
The coupling parameter $g_{p,\nu; p',\nu}$ (taken as a constant $g_{\nu}$ in simulations)
relates the NESB model to the boson picture via the relation \cite{legget}
\bea
\alpha_{\nu}=\frac{1}{2}\left[ \frac{2}{\pi} {\rm atan} ( \pi\rho_{\nu}(\epsilon_F)g_{\nu}) \right]^2.
\label{eq:galph}
\eea
Here $\rho_{\nu}(\epsilon_F)$ is the density of states at the Fermi energy
and $\alpha_{\nu}$ is a dimensionless parameter, the prefactor in the bosonic-Ohmic spectral function,
see Eq. (\ref{eq:spect}).

We had recently simulated the energy current characteristics of the unbiased NESF model
by adapting the INFPI approach  \cite{segal13}.
We discuss next the structure of the current operator in this simulation. First, we transform the 
Hamiltonian (\ref{eq:SF}) via a unitary transformation $\tilde H_{SF}=U^{\dagger}H_{SF}U$ as in Sec.
\ref{BRHam} into
\bea
\tilde H_{SF}=H_S+H_F+V,
\eea
where 
\bea
H_S&=&\frac{\hbar\epsilon}{2} \tilde \sigma_z,
\nonumber\\
H_F&=&H_L+H_R, \,\,\,\,\,\, H_{\nu}=\sum_{p}\epsilon_p
c_{p,\nu}^{\dagger} c_{p,\nu},
\nonumber\\
V&=&V_L+V_R, \,\,\,\,\,\,\,
V_{\nu}=\tilde\sigma_x\sum_{p,p'}\hbar g_{p,\nu; p' \nu}c_{p,\nu}^{\dagger}c_{p',\nu }
\label{eq:H}
\eea
As before, $\tilde \sigma_{x,y,z}$ stand for Pauli matrices in the energy basis.
We assume a factorized initial state, $\tilde\rho(0)=\tilde\rho_S(0) \otimes \rho_L\otimes\rho_R $,
$\tilde\rho_S$ denotes the
reduced density matrix of the subsystem and
$\rho_{\nu}=e^{-\beta_{\nu}(H_{\nu}-\mu_{\nu}N_{\nu})}/{\rm
Tr_{\nu}}[e^{-\beta_{\nu}(H_{\nu}-\mu_{\nu}N_{\nu})}]$.
In our simulations below we take $\mu_L=\mu_R$ but assume different
(time-zero) temperatures for the thermal baths, $T_L\neq T_R$.
At $t=0$ we put into contact the two Fermi baths through the
quantum subsystem, then follow the evolution of the reduced density
matrix and the energy current until (quasi) steady-state sets in.
Since electron flow is blocked and energy is transferred
through excitation - de-excitation processes of the TLS,
we refer to the energy current here as a ``heat current".

The heat current in the unbiased ($\omega_0=0$)  model was calculated in the energy
basis  using the construction \cite{claire09}
\bea
 j_L = {\rm Tr}_S{\rm Tr}_F[\tilde {\hat j}_L\rho(t)] =
-\frac{i}{\hbar} {\rm Tr}_S{\rm Tr}_F \{\tilde \rho(t) [H_S,V_L]\},
\label{eq:J2}
\eea
obtained from the definition (\ref{eq:defjq}) under a  steady-state assumption.
Here, ${\rm Tr}_F$ (${\rm Tr}_S$)
refers to a partial trace over the Fermi-sea electrons (TLS). 
We identify the current with the contact at which it is evaluated, though $j_q=j_L=-j_R$ is satisfied here.
The commutator can be readily performed to yield
\bea
[H_S,V_L]=i\hbar^2 \Delta\tilde\sigma_y\sum_{l,l'}g_{l,L;l',L}c_{l,L}^{\dagger}c_{l',L},
\eea
leading to
\bea
j_L=
(\hbar\Delta) {\rm Tr}_S\left[ \tilde\sigma_y{\rm Tr_F}[A_L\tilde\rho(t)]\right].
\eea
The bath operator is given by $A_L\equiv \sum_{l,l'} g_{l,L;l',L}c_{l,L}^{\dagger}c_{l',L}$.
We now further define a subsystem operator as
\bea A_S(t)\equiv{\rm Tr_F}[A_L\rho(t)]= {\rm
Tr_F}[e^{iHt/\hbar}A_Le^{-iHt/\hbar}\tilde\rho(0)], \label{eq:Ast}
\eea
and express the current with its matrix elements
\bea j_L=(\hbar\Delta) [-i (A_S(t))_{-,+} + i
(A_S(t))_{+,-} ]. \label{eq:curr}
\eea
We use INFPI to time-evolve Eq. (\ref{eq:Ast}) \cite{segal13}. These Simulations are
compared to BR and NIBA results from Secs. \ref{BRH} and \ref{NIBAC}, respectively.



\begin{figure}[htbp]
\vspace{0mm} \hspace{0mm}
{\hbox{\epsfxsize=70mm \epsffile{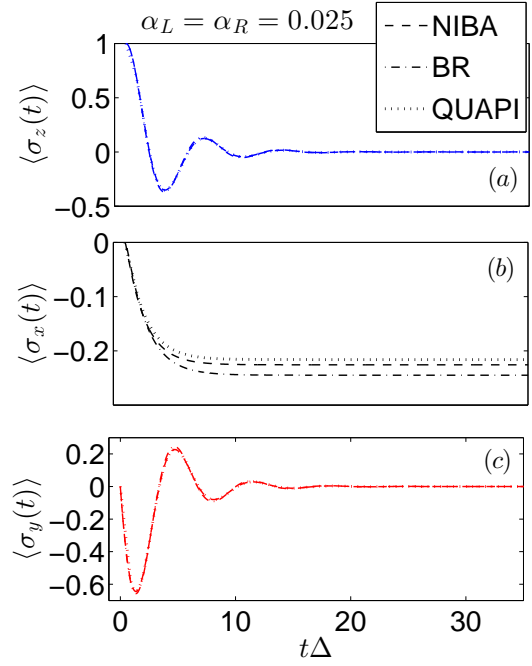}}}
\caption{
Dynamics of the spin subsystem in the NESB model
assuming an Ohmic spectral function with
$\omega_c=20\Delta$, and $k_BT_L=k_BT_R=2\hbar\Delta$, $\omega_0=0$, $\alpha_L=\alpha_R=0.025$.
(a-c) Results for $\langle\sigma_{x,y,z}(t)\rangle$ are presented in the local basis ($|0\rangle$ and $|1\rangle$),
 QUAPI (dotted), NIBA (dashed), BR (dashed-dotted).
}
\label{Fig1}
\end{figure}

\begin{figure}[htbp]
\vspace{0mm} \hspace{0mm}
{\hbox{\epsfxsize=70mm \epsffile{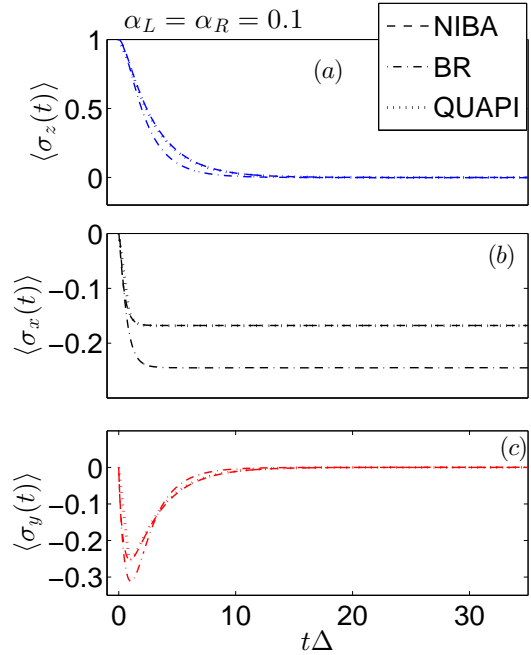}}}
\caption{
Dynamics of the spin subsystem with the parameters of Fig. \ref{Fig1},
$\alpha_L=\alpha_R=0.1$.
}
\label{Fig2}
\end{figure}

\begin{figure}[htbp]
\vspace{0mm} \hspace{0mm}
{\hbox{\epsfxsize=70mm \epsffile{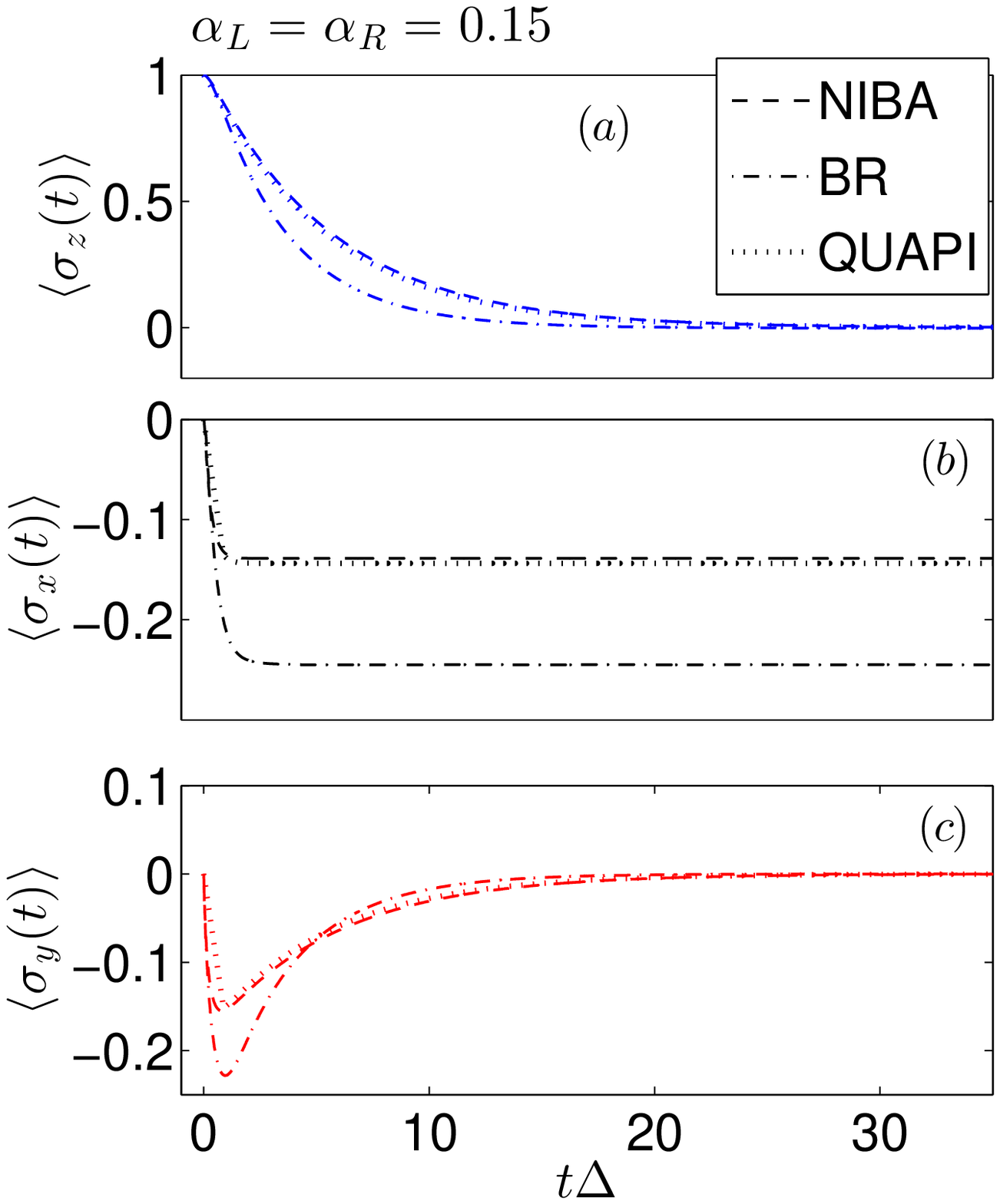}}}
\caption{
Dynamics of the spin subsystem as in Fig. \ref{Fig1}, with
$\alpha_L=\alpha_R=0.15$.
}
\label{Fig3}
\end{figure}


\section{Simulations}
\label{results}

The dynamics of the reduced density matrix is obtained by solving numerically
the BR and NIBA integro-differential equations, Eq. (\ref{eq:BRdyn}) and (\ref{eq:NIBAintegro}), respectively,
using the trapezoidal rule for the inner integral, see Ref. \cite{integro}.
This brute-force approach was adopted with a small time step, $\delta t=T_p/2000$; the period is defined by $T_p\equiv2\pi/\epsilon$.
Integrals over frequency were evaluated using the trapezoidal rule, discretized with
$\delta \omega\sim\epsilon/2000$, up to the limit $30\omega_c$.
QUAPI and INFPI simulations were performed with $\delta t\sim T_p/50$.
Other parameters are $k_BT/\hbar\Delta\sim 1-2$,
$\omega_0/\Delta=0-5$, $\omega_c=20\Delta$ and $\alpha_{\nu}=10^{-3}-1$.
Thermal conductances were calculated by taking a small temperature difference,
$2(T_L-T_R)/[(T_L+T_R)]=0.05$. INFPI simulations better converge for larger temperature differences,
$k_B(T_L-T_R)/\hbar\Delta\sim1$.
We use here an Ohmic spectral function, but we are not limited to this form and
other environments can be similarly explored, e.g., a Debye spectral function
or a spin bath mimicked by a harmonic environment, with a temperature-dependent spectral function \cite{Makriproof}.

\subsection{Unbiased model}

{\it RDM Dynamics.}
We begin with the unbiased $\omega_0=0$ model. In Figs. \ref{Fig1}-\ref{Fig3}
we follow the dynamics  of the subsystem in the local basis $(|0\rangle$, $|1\rangle$)
using BR, NIBA and QUAPI  \cite{QUAPI}, extended here to the non-equilibrium two-bath case.
At very weak coupling, $\alpha_{\nu}=0.025$, BR and NIBA reasonably agree with QUAPI;
at smaller couplings, $\alpha<0.02$, the agreement is excellent (not shown).
Increasing the coupling to $\alpha=0.1-0.15$, translating to a
total interaction strength $\alpha=0.2-0.3$,
we note that NIBA and BR agree with QUAPI over
the polarization dynamics,
but the real part of off-diagonal elements deviate up to a factor of 1.5 in the steady-state limit.

\begin{figure}[htbp]
\vspace{0mm} \hspace{0mm}
{\hbox{\epsfxsize=90mm \epsffile{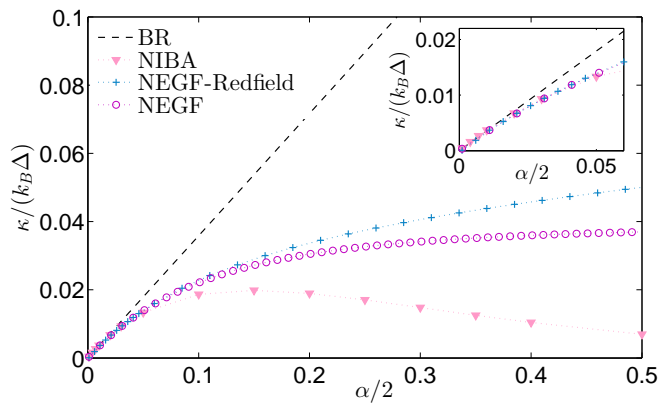}}}
\caption{
Thermal conductance in the unbiased NESB model as a function of $\alpha/2=\alpha_{L}=\alpha_R$.
We use Ohmic spectral functions with
$\omega_c=20\Delta$,  $\omega_0=0$, $k_BT_L=2.05\hbar\Delta$,
$k_BT_R=1.95\hbar\Delta$ with
BR (dashed), NIBA ($\triangledown$), NEGF-Redfield from Ref. \cite{vtw10} ($+$)
and the NEGF expression
(\ref{eq:Yang}) from Ref. \cite{wu14} ($\circ$).
The inset zooms over the weak coupling regime.}
\label{Fig4}
\end{figure}

{\it Heat transfer: Coupling to the contacts.}
Deviations between BR and NIBA in the RDM behavior propagate into significant, qualitative differences
in the heat current characteristics.
In Fig. \ref{Fig4} we display the thermal conductance of the NESB model and show that within the BR scheme it grows linearly with $\alpha$ \cite{sn05}, while NIBA  demonstrates saturation of the current, then its decay
at large values of $\alpha$ \cite{segal14}.
We also illustrate the behavior of the thermal conductance using the NEGF-Redfield scheme
of Ref. \cite{vtw10}. In this approach correlation functions in the (exact)
heat transfer Meir-Wingreen formula are evaluated at the level of the Redfield theory, which is second order in
the system-bath coupling. A different NEGF-based expression for the heat current was proposed in Ref. \cite{wu14}.
It was obtained from the diagrammatic expansion, by calculating the
spin-spin correlation functions via the Majorana-fermion representation of spin operators, truncating diagrams of high order system-bath couplings.
This results in an elegant expression (presented here for $\omega_0=0$),
\bea
j_q=\frac{2}{\pi}\int_0^{\infty}\frac{\hbar\omega \Delta^2\Gamma_R(\omega)\Gamma_L(\omega)}{(\omega^2-\Delta^2)^2+[\Gamma_L(\omega)\coth(\beta_L\hbar\omega/2)+\Gamma_R(\omega)\coth(\beta_R\hbar\omega/2)]^2\omega^2)}[n_L(\omega)-n_R(\omega)]d\omega.
\label{eq:Yang}
\eea
This formula resembles the harmonic limit \cite{Har}, and it nicely interpolates between the
very weak coupling equation (\ref{eq:jweak}), resulting from a kinetic-type dynamics, 
and the low-temperature limit, where heat is transferred coherently, 
out of resonance with the central frequency; the integration over frequencies away from $\Delta$ 
reflects a tunneling behavior.

We see in Fig. \ref{Fig4} that the two NEGF-based approaches perform very well up to $\alpha_{\nu}=0.1$.
Predictions beyond that are qualitatively incorrect as these methods fail to provide
the decay behavior of the current with $\alpha$ at strong coupling.
However, even in this regime NEGF-Redfield and the NEGF expression (\ref{eq:Yang}) are valuable tools;
their predictions are significantly closer to NIBA than to BR.

\begin{figure}[htbp]
\vspace{0mm} \hspace{0mm}
{\hbox{\epsfxsize=90mm \epsffile{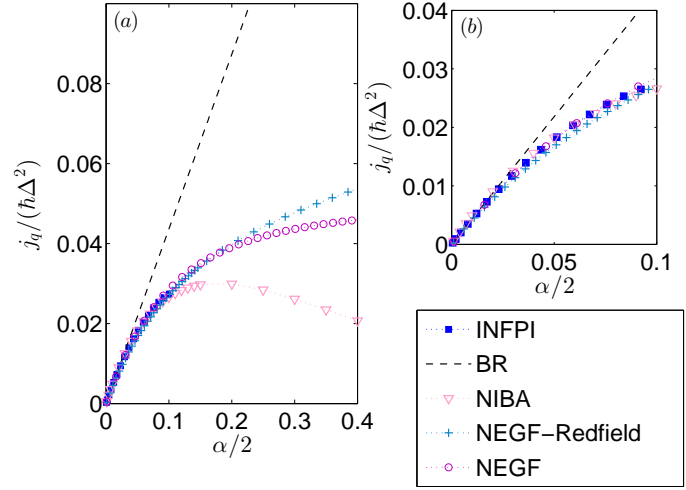}}}
\caption{ (a)
Heat current in the unbiased NESF model. Simulations were performed in the fermionic picture using
electron bands with linear dispersion and a hard cutoff at  $D/\Delta=\pm 5$.
The parameters of the fermionic model are matched with the bosonic picture using
Eq. (\ref{eq:galph}),
$\alpha_L=\alpha_R$, $\omega_0=0$, $k_BT_L=2\hbar\Delta$,
$k_BT_R= \hbar\Delta$.
We compare
INFPI ($\square$) 
to simulations in the bosonic picture,
BR (dashed), NIBA ($\triangledown$), NEGF-Redfield  \cite{vtw10} ($+$) and the NEGF expression
(\ref{eq:Yang}) \cite{wu14} ($\circ$).
Panel (b) zooms over the small-$\alpha$ regime.
INFPI numerical parameters are $\delta t=0.1/\Delta$ and $N_s=9$, for more details see Ref. \cite{segal13}.
}
\label{FigSF}
\end{figure}

In Fig. \ref{FigSF} we use INFPI to simulate the
heat current in the NESF model \cite{segal13}. We display the
current as a function of the interaction parameter $\alpha$ up to $\alpha/2=0.1$, 
which we consider as a weak-intermediate value. 
We compare INFPI results to the predictions of BR, NIBA, NEGF-Redfield \cite{vtw10} and NEGF of Ref. \cite{wu14}.
BR equation fails beyond $\alpha/2=0.02$, but the other methods excellently agree. 
Note that in the fermionic language $\alpha/2=0.1$ translates to a phase shift of
$\phi_{\nu}=\pi\rho_{\nu}\alpha_{\nu}\sim 1$.
It is useful to comment at this point that a fully harmonic junction is expected to provide higher currents
than supported by the (anharmonic) NESB junction \cite{Har}. Specifically, under a weak coupling approximation
the current in harmonic junctions obeys Eq.
(\ref{eq:jweak}), only missing the temperature-dependent distribution functions in the denominator.

We now explain our classification of system-bath coupling domains
based on Figs. \ref{Fig4}-\ref{FigSF}.
In the very weak coupling limit $\alpha_{\nu}<0.025$
the Bloch-Redfield treatment is valid within up to $10\%$ deviations from exact results.
In the so-called weak coupling regime $0.025<\alpha_{\nu}<0.1$
 deviations from linearity are apparent, but the current is growing monotonically
with $\alpha$. For stronger couplings,  $0.1<\alpha_{\nu}<0.5$, the heat current displays a crossover behavior:
Fig. \ref{Fig4} shows that beyond $\alpha_{\nu}=0.15$ the thermal conductance drops
with increasing couplings to the bath.
We refer to this crossover area as the intermediate coupling regime.
At even stronger coupling the behavior of $j_q(\alpha)$, or the thermal conductance, is dominated by
an exponentially decaying factor of $\alpha$, see Eq. (\ref{eq:kapNiba}).

\begin{figure}[htbp]
\vspace{0mm} \hspace{0mm} {\hbox{\epsfxsize=70mm
\epsffile{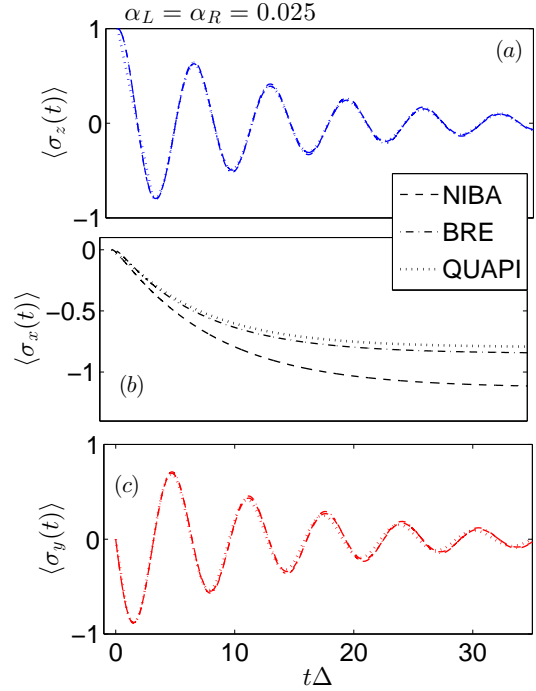}}} \caption{Dynamics of the unbiased model beyond the non-adiabatic region
for $\omega_c=4\Delta$, $k_BT_L=k_BT_R=\hbar\Delta/2.5$, and
$\alpha_L=\alpha_R=0.025$. QUAPI (dotted), NIBA (dashed),
BR (dashed-dotted).
} \label{dynDel}
\end{figure}

{\it Beyond the non-adiabatic limit.}
The non-adiabatic limit describes a junction comprising a low-frequency (slow) central vibration,
relative to the cutoff frequency of the bath.
In Fig. \ref{dynDel} we abandon the strict non-adiabatic region and explore the dynamics when 
$\Delta/\omega_c=0.25$
with $2.5k_BT_{\nu}=\hbar\Delta$.
 We study the time evolution of the RDM at very weak coupling and
find that NIBA fails to reproduce $\langle\sigma_x(t)\rangle$, deviating by $\sim 40\%$, while it
performs well for the polarization dynamics. We now discuss implications on transport properties.

\begin{figure}[htbp]
\vspace{0mm} \hspace{0mm}
{\hbox{\epsfxsize=90mm \epsffile{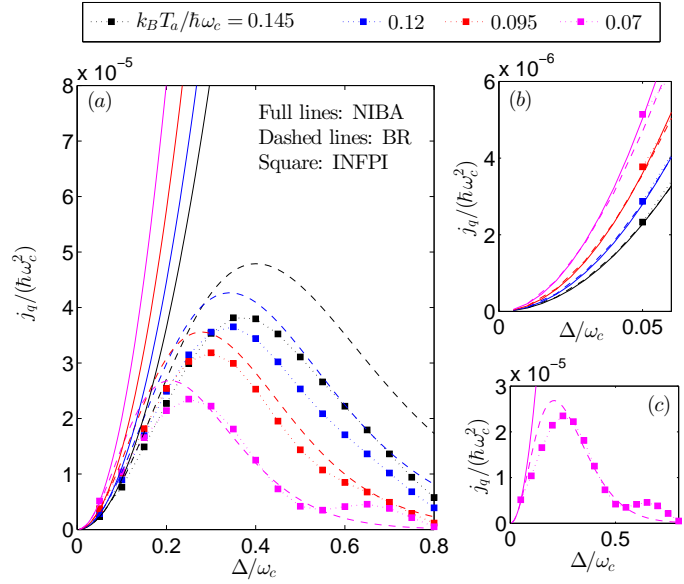}}}
\caption{
Heat current as a function of the TLS tunneling frequency.
(a) INFPI simulations in the fermionic picture ($\square$).
The parameters of the fermionic model are matched with the bosonic case using
Eq. (\ref{eq:galph}),
$\pi\rho(\epsilon_F)g=0.3$, leading to
$\alpha_L=\alpha_R=0.0172$. Other parameters are $\omega_0=0$,
$k_BT_a/\hbar\omega_c$ as indicated in the legend with $T_a=(T_L+T_R)/2$
and $(T_L-T_R)/\omega_c=0.01$.
We compare INFPI ($\square$) results to simulations in the bosonic picture,
BR (dashed), NIBA (full). 
Panel (b) zooms over the non-adiabatic regime.
INFPI numerical parameters are $\delta t=0.1/\Delta$ and $N_s=9$, for more details see Ref. \cite{segal13}.
Panel (c) zooms over the low temperature case,  $k_BT_a/\hbar\omega_c=0.07$.
}
\label{FigSFDEL}
\end{figure}

The effect of the spin tunneling frequency $\Delta$
on the steady-state heat current is displayed in Fig. \ref{FigSFDEL}, considering a weakly coupled nanojunction.
The heat current exhibits a turnover behavior as a function of $\Delta$, explained based on the BR expression (\ref{eq:jweak}): The current
increases with $\Delta$, the quanta transmitted,
but it further requires a sufficient thermal occupation factor in matching bath modes.
Thus, when the tunneling frequency is high,
$\Delta >2k_BT/\hbar$, the current begins to drop due to reduced population of in-resonance  modes.
Comparing BR (dashed) to INFPI (square), we find that the BR scheme
reproduces the correct turnover behavior, though the exact position of the maxima
is shifted, and the magnitude of the current is overestimated at high temperatures \cite{commSF}.
NEGF results are not displayed here; in this weak-coupling regime they overlap with BR.
In contrast, NIBA equations miss altogether the correct behavior of $j_q(\Delta)$ once $\Delta/\omega_c>0.1$
since NIBA only captures non-adiabatic contributions, $j_q\propto \Delta^2$.
Moreover, it is interesting to note that not only does NIBA provide the wrong functional form for $j_q(\Delta)$,
it further predicts an erroneous temperature-dependent behavior:
While in the non-adiabatic regime at small coupling the current drops with increasing temperatures,
$j_q\propto T^{-1}$,
a tendency captured by NIBA, the opposite behavior takes place beyond the non-adiabatic regime
once the current is controlled by thermal occupation factors in the contacts.

Another interesting observation concerns the low temperature $k_BT_a/\hbar\omega_c=0.07$ regime, see Fig. \ref{FigSFDEL}(c).
In this case INFPI simulations indicate on the existence of {\it two}
peaks in the current $j_q(\Delta)$, at $\Delta/\omega_c$=0.25,  0.65.
The first peak corresponds to
absorption and emission processes of a single phonon in either the $L$ or $R$ baths, as explained above.
In the language of the SF model, these are single electron-hole pair generation or destruction processes.
Around $\Delta/\omega_c$=0.25,  the enhancement of the current with an enlarging spin frequency
is balanced by thermal occupation factors of bath modes.
We presume that the second peak corresponds to a similar balance, apparently reflecting
two-phonon processes (two electron-hole pairs) participating in the
excitation and relaxation dynamics of the TLS. Additional simulations are required to establish this result.

\begin{figure}[htbp]
\vspace{0mm} \hspace{0mm}
{\hbox{\epsfxsize=90mm \epsffile{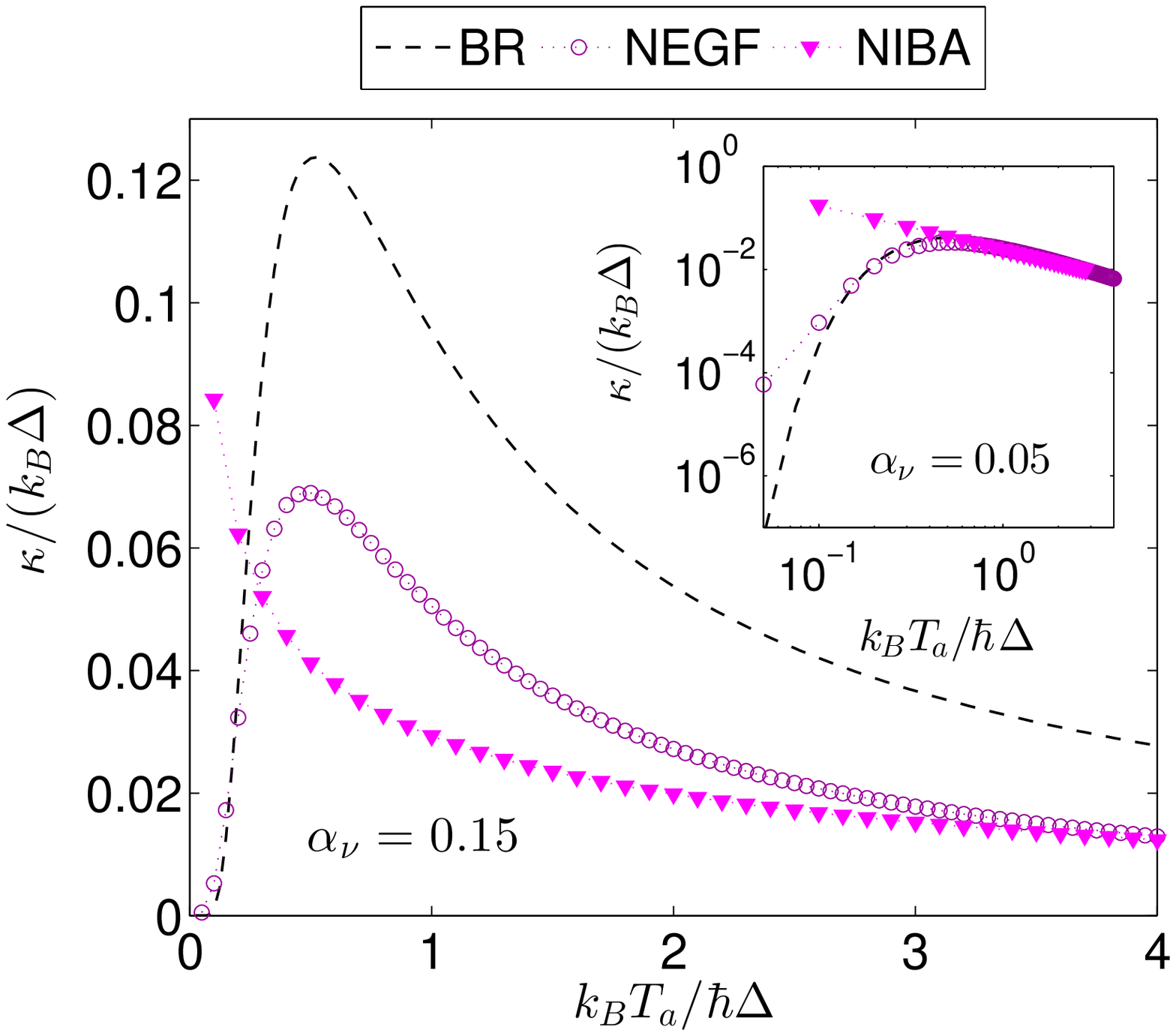}}}
\caption{
Thermal conductance in the unbiased NESB model as a function of temperature.
We use Ohmic spectral functions with
$\omega_c=20\Delta$, $\alpha_{\nu}=0.15$.
BR (dashed), NIBA ($\triangledown$), 
and the NEGF expression (\ref{eq:Yang}) \cite{wu14} ($\circ$).
NEGF-Redfield formula did not provide physical answers at high temperatures here.
The inset displays a weak coupling example with $\alpha_{\nu}=0.05$.}
\label{FigT}
\end{figure}

{\it Heat transfer: Temperature dependence.}
The temperature dependence of the thermal conductance is of particular interest for actual devices;
exact Monte-Carlo simulations \cite{saitoMC}
provided the high temperature behavior of the NESB model
$\kappa \propto \left(\frac{k_B T}{\hbar \omega_c}\right)^{2\alpha-1}$
and the low-temperature scaling
$\kappa \propto \alpha  (T/T_K)^3$,
$T_K$ is the Kondo temperature in the system,
a function of the microscopic parameters $\Delta$, $\omega_c$ and $\alpha$.
We had recently proved that the high-temperature limit is reproduced by NIBA-heat current formula, see
Eq. (\ref{eq:kapNiba}) \cite{segal14}.
Fig. \ref{FigT} displays the thermal conductance characteristics with temperature in the non-adiabatic limit
at intermediate and weak (inset) coupling.
In the weak coupling limit the three techniques coincide at high temperatures.
In contrast, at stronger coupling the (correct) NIBA scaling is neither reproduced by BR nor by NEGF.
It is also significant to comment that NIBA misses altogether the enhancement of the conductance with $T$
for $k_BT_a/\hbar \Delta < 0.5$. Therefore, it is crucial to carefully examine parameters of interest
to determine which technique is most suitable.


\subsection{Biased model}

\begin{figure}[htbp]
\vspace{0mm} \hspace{0mm} {\hbox{\epsfxsize=70mm
\epsffile{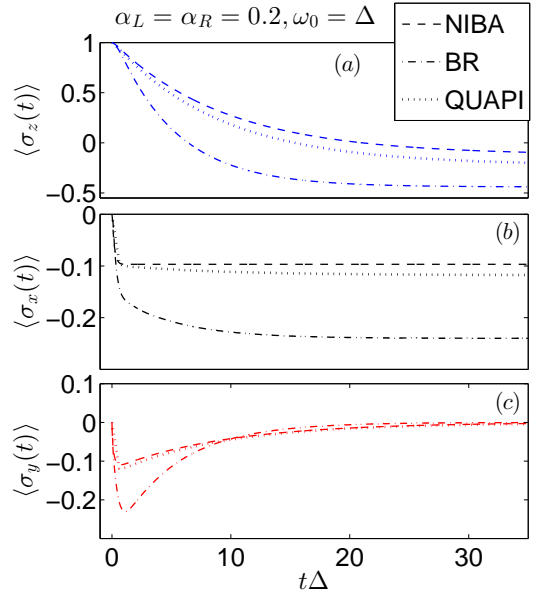}}} \caption{Dynamics of the biased model
$\omega_0=\Delta$ in the intermediate coupling regime,
$\alpha_L=\alpha_R=0.2$, and $\omega_c=20\Delta$,
$k_BT_L=k_BT_R=2\hbar\Delta$. QUAPI (dotted), NIBA (dashed),
BR (dashed-dotted).
} \label{Fig6}
\end{figure}

{\it RDM dynamics.}
In Fig. \ref{Fig6} we display the RDM dynamics in the biased model. At weak coupling NIBA
misses the correct behavior of the RDM \cite{weiss}
while at strong coupling it reasonably agrees with QUAPI.

\begin{figure}[htbp]
\vspace{0mm} \hspace{0mm}
{\hbox{\epsfxsize=70mm \epsffile{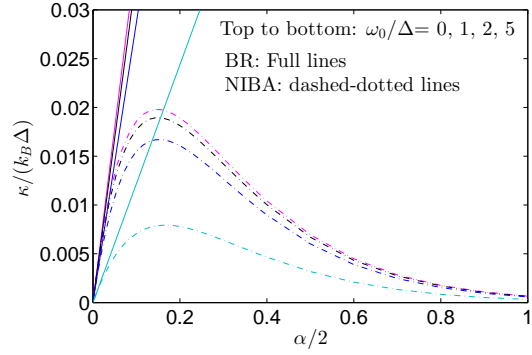}}}
\caption{
Thermal conductance for the biased model
assuming Ohmic spectral functions and $\alpha_L=\alpha_R$,
$\omega_c=20\Delta$, $k_BT_L=2.05\hbar\Delta$,
$k_BT_R=1.95\hbar\Delta$.
}
\label{Fig7}
\end{figure}

{\it Thermal conductance.}
Since the NIBA-heat current expression
only depends on the polarization dynamics (steady-state value and the relaxation rates),
errors in $\langle\sigma_x(t)\rangle$ do not propagate into the calculation of the heat current,
 see Fig. \ref{Fig7}.
The bias affects the conductance in a simple way: In the BR scheme
$\kappa \propto \epsilon/\sinh(\beta\hbar\epsilon)$, see Eq. (\ref{eq:BRkappa2}).
 In NIBA calculations we confirmed numerically
that $\kappa \propto \omega_0/\sinh(\beta\hbar\omega_0)$, for $\alpha=0.1-0.5$.
Thus, the bias does not offer a new ``quantum" control knob over the heat current
as $\omega_0$ does not tangle with the coupling strength $\alpha$ in the present non-adiabatic limit.

\section{Summary}
\label{Summ}

We provided a comprehensive analysis of the non-equilibrium spin-boson
model, comprising a spin subsystem coupled to two thermal baths of different temperatures.
We studied the dynamics of the spin reduced density matrix and the
transfer of heat in the model using different techniques: the Bloch-Redfield scheme which is
valid in the very weak system-bath
coupling limit, the noninteracting-blip approximation, 
exact in the non-adiabatic ($\Delta\ll\omega_c$) scaling limit,
and numerically-exact influence functional path integral simulations.
We also compared results to NEGF-based techniques \cite{vtw10,wu14}.
The BR, NIBA and path integral approaches were originally
developed for exploring decoherence effects and dissipation in open quantum systems. Here we bring
an organized discussion of their extensions to treat transport behavior.

Specific observations include: (i) The biased and unbiased NESB models may
display similar dynamics within the BR and NIBA approaches,
but relatively small deviations in the RDM behavior propagate
into strong and qualitative disagreements in the heat current characteristics.
(ii) The BR formalism should be used with great caution in
modeling actual devices since the prediction  $j_q \propto \alpha$ fails beyond the very weak coupling limit,
providing unphysical-incorrect  large conductances.
(iii)
In the regime of validity for BR and NIBA,  the spin bias (detuning) parameter
does not offer a new-nontrivial control mean over the heat current; at large detuning the current decays since
 thermal occupation of high frequency bath modes (above the thermal energy) is reduced.
(iv) In the non-adiabatic regime at weak coupling and high temperatures, the thermal conductance
decreases with increasing temperatures, $\kappa\propto (T_L-T_R)/T$.
This behavior stems from the intrinsic anharmonicity of the junction.
This trend is correctly captured by the BR method, NIBA and INFPI.
In contrast, beyond the non-adiabatic limit this trend is reversed since 
 thermal occupation factors of bath modes dominate the current
rather than temperature-dependent (anharmonic) scatterings
in the junction. As expected, NIBA fails to capture this behavior:
It overestimates the current by orders  of
magnitude and it predicts an enhancement of the current when reducing the temperature,
irrespective of the frequency $\Delta$.

It is of interest to extend our analysis and further explore the behavior under 
classical equations of motion, or mixed quantum-classical treatments \cite{mixed}.
Future studies will also examine the time-dependent-driven NESB model
\cite{grifoni,jie10,jie13,uchi} with the objective to understand
the role of strong system-bath couplings and quantum coherence in possibly
enhancing heat to work conversion efficiency.

\begin{acknowledgments}
This work was funded by the Natural Sciences and Engineering
Research Council of Canada and the Canada Research Chair Program. NB
acknowledges support from the  CQIQC Undergraduate Summer Research
Studentships.
\end{acknowledgments}


\begin{thebibliography}{99}
\bibitem{Har}
The heat current through a single mode (frequency $\Delta$) harmonic junction is given by
$j_q=\frac{2}{\pi}\int_0^{\infty} \hbar\omega\mathcal T(\omega) [n_L(\omega)-n_R(\omega)]d\omega$
with $\mathcal T(\omega)=\frac{\omega^2\Gamma_L(\omega)\Gamma_R(\omega)}{(\omega^2-\Delta^2)^2+[\Gamma_R(\omega)+\Gamma_L(\omega)]\omega^2}$
\cite{segal07}.
\bibitem{commSF}
 BR results of Fig. \ref{FigSFDEL} were calculated in the fermionic picture, by following Ref. \cite{segal13}
with the Golden-Rule transition rates evaluated by an explicit integration, to allow the band-edge 
effects to be properly accounted for. Cutoff effects become more pornounced at high temperaures and large 
tunneling frequencies, $k_BT/\hbar\omega_c>0.2$, $\Delta/\omega_c>0.1$.
\end{thebibliography}
\end{document}